\documentclass[12pt]{article}
\pdfoutput=1 

\usepackage{cancel,slashed}
\usepackage{bbm}
\usepackage{epsf}
\usepackage{amssymb}
\usepackage{amsfonts}
\usepackage{amsmath}
\usepackage{graphicx}
\usepackage{cite}

\usepackage{latexsym}
\usepackage{color}
\usepackage{xypic}
\usepackage{cancel,slashed}
\usepackage[linktocpage]{hyperref}
\usepackage{color}
\usepackage{transparent}

\newcommand{\bmat}{\left(\begin{array}}
\newcommand{\emat}{\end{array}\right)}

\def\p{\partial}
\def\a{\alpha}

\def\b{\beta}
\def\g{\gamma}
\def\d{\delta}
\def\th{\theta}

\def\-{\hphantom{-}}

\def\s2{\frac{1}{\sqrt2}}

\def\oh{\frac{1}{2}}
\def\beq{\begin{equation}}
\def\eeq{\end{equation}}
\def\beqa{\begin{eqnarray}}
\def\eeqa{\end{eqnarray}}

\def\im{{\rm Im \,}}
\def\re{{\rm Re \,}}

\def\Dsl{\,\raise.15ex\hbox{/}\mkern-13.5mu D} %this one can be subscripted

\def\CM {{\cal M}}

\def\CN {{\cal N}}
\def\CF {{\cal F}}

\def\CV {{\cal V}}
\def\CW {{\cal W}}

\def\CJ {{\cal J}}
\def\re{\mbox{Re}}
\def\im{\mbox{Im}}

\def\be{\begin{equation}}
\def\ee{\end{equation}}
\def\bea{\begin{eqnarray}}
\def\eea{\end{eqnarray}}
\def\raw{\rightarrow}

\def\bes{\begin{subequations}}
\def\ees{\end{subequations}}

\def\IZ{\mathbb{Z}}
\def\IR{\mathbb{R}}

\def\oh{\frac{1}{2}}
\def\a{{\alpha}}
\def\b{{\beta}}
\def\d{{\delta}}

\def\th{{\theta}}

\def\lam{{\lambda}}

\def\g{{\gamma}}

\def\p{{\partial}}

\def\dj{{d^{\mathcal J}}}

\def\sm2{{\mbox{\small 2}}}

\begin{document}
\pagestyle{plain}

%----------------------------------------------------------------------%
%  numbering equations with section number
%----------------------------------------------------------------------%
\makeatletter
\@addtoreset{equation}{section}
\makeatother
\renewcommand{\theequation}{\thesection.\arabic{equation}}
%----------------------------------------------------------------------%
%  title page
%----------------------------------------------------------------------%
\pagestyle{empty}
%\vspace*{1.0in}
\rightline{IFT-UAM/CSIC-14-098}
%\rightline{\tt hep-th/yymmnnn}
\vspace{0.5cm}
\begin{center}
\LARGE{{On D-brane moduli stabilisation}
\\[10mm]}
\large{Fernando Marchesano,$^1$ Diego Regalado$^{1,2}$ and Gianluca Zoccarato$^{1,2}$ \\[10mm]}
\small{
${}^1$ Instituto de F\'{\i}sica Te\'orica UAM-CSIC, Cantoblanco, 28049 Madrid, Spain \\[2mm] 
${}^2$ Departamento de F\'{\i}sica Te\'orica, 
Universidad Aut\'onoma de Madrid, %Cantoblanco, 
28049 Madrid, Spain
\\[20mm]} 
\small{\bf Abstract} \\[5mm]
\end{center}
\begin{center}
\begin{minipage}[h]{15.0cm}

Standard results in 4d $\CN=1$ string compactifications assign a number of moduli to each space-time filling D-brane, computed by analysing the D-brane action in a fixed background. We revisit such conventional wisdom and argue that this naive counting of open string moduli is incorrect, in the sense that some of them will be lifted when making dynamical the bulk degrees of freedom. We explicitly discuss this effect for D6-branes wrapping special Lagrangian three-cycles, showing that some geometric and Wilson line moduli are lifted even before taking into account worldsheet instanton effects. From a 4d effective theory viewpoint the moduli lifting is due to an F-term potential, and can be deduced from the superpotentials in the literature. From a microscopic viewpoint the lifting is due to D-brane backreaction effects and flux quantisation in a compact manifold, and provides a mechanism for lifting Wilson line moduli. The latter applies to certain D6-branes and D7-brane Wilson lines, yielding new possibilities to build models of inflation in string theory. 

\end{minipage}
\end{center}
\newpage
%----------------------------------------------------------------------%
%  Resetting of counters
%----------------------------------------------------------------------%
\setcounter{page}{1}
\pagestyle{plain}
\renewcommand{\thefootnote}{\arabic{footnote}}
\setcounter{footnote}{0}
%----------------------------------------------------------------------%
%  Paper begins
%----------------------------------------------------------------------%

%\end{document}

\tableofcontents

%\newpage

\section{Introduction}
\label{s:intro}

One of the most generic features of standard 4d string vacua is the presence of several massless and/or very light neutral scalar fields, whose presence prevents to construct realistic string models of Particle Physics and Cosmology \cite{thebook,reviews}. Such moduli problem has been the subject of intense research over the years, with the result of several scenarios in which moduli in the gravity sector are stabilised by a combination of background fluxes and possibly non-perturbative effects \cite{Grana:2005jc,Douglas:2006es,Denef:2007pq}. 

This encouraging picture has been mostly developed for the closed string sector of type II string theories. Nevertheless, to construct a realistic model one should also check that there are no unwanted moduli also in the gauge sector of the compactification. In particular, in the context of type II vacua, one should check that there are no D-brane moduli preventing a realistic particle spectrum. 

The computation of open string moduli is typically done by considering a probe BPS D-brane in a supersymmetric closed string background and then counting all the possible D-brane deformations that preserve its BPS conditions. For instance, one may consider a closed string background of the form $\IR^{1,3} \times \CM_6$, where $\CM_6$ is a Calabi-Yau three-fold, and a single D$p$-brane on $\IR^{1,3} \times \Pi_{p-3}$, with $\Pi_{p-3}$ a compact cycle of $\CM_6$. Then, by either a worldsheet \cite{Ooguri:1996ck,Kapustin:2003se} or a D-brane worldvolume \cite{Becker:1995kb,Bergshoeff:1997kr,Marino:1999af} analysis, one determines the conditions that $\Pi_{p-3}$ must satisfy to preserve some of the supersymmetry of the background. Finally, one studies the deformations of the internal D-brane embedding $\Pi_{p-3}$ and of the U(1) bundle on it that preserve such supersymmetry conditions, in order to determine the number of massless deformation or moduli of such D-brane.

\begin{table}[htb] 
%\footnotesize
\renewcommand{\arraystretch}{1.2}
\setlength{\tabcolsep}{5pt}
\begin{center}
\begin{tabular}{|c|c|c|}
\hline
& D6-brane & D7-brane \\
\hline\hline
 BPS condition & $\Pi_3$ special Lagrangian & $\Pi_4$ holomorphic \\
\hline
complex moduli & $b_1(\Pi_3)$ & $h^{1,0}(\Pi_4) + h^{2,0}(\Pi_4)$\\
\hline
\end{tabular}
\end{center}
\caption{BPS conditions and moduli for D6/D7-branes with flat bundles in a Calabi-Yau. $h^{n,0}(\Pi_4)$ counts harmonic ($n,0$)-forms on $\Pi_4$ and $b_1(\Pi_3)$ counts harmonic one-forms of $\Pi_3$.}
\label{t:BPS}
\end{table}

Table \ref{t:BPS} shows the result of such analysis for the simple case of D6 and D7-branes with flat U(1) bundles in a Calabi-Yau three-fold $\CM_6$.\footnote{More precisely, one should consider such space-time filling D-branes in a Calabi-Yau modded out by an orientifold involution. Implementing such orientifold projection will affect the definition of the D-brane fields and in particular of its moduli, see \cite{Jockers:2004yj,gl11,kt11} for careful analysis. This subtlety will however not be relevant for our line of reasoning and we may ignore it for the purposes of the present discussion.} For D7-branes, $h^{2,0}(\Pi_4)$ corresponds to the independent geometric deformations of $\Pi_4$ that preserve its holomorphicity, while $h^{1,0}(\Pi_4)$ counts the complex flat U(1) bundles or Wilson lines that can be turned on. For D6-branes, the number of deformations of $\Pi_3$ that preserve the special Lagrangian condition is given by  $b_1(\Pi_3)$ \cite{McLean}%, that is the number of independent harmonic one-forms on $\Pi_3$ or equivalently the number of independent two-cycles $\pi_2 \subset \Pi_3$
. Finally, each D6-brane position is completed into a complex scalar by a D6-brane real Wilson line, whose number is also counted by $b_1(\Pi_3)$.

This moduli counting can be partially modified for more involved compactifications. Indeed, for the case of the D7-branes one may consider endowing them with non-trivial worldvolume fluxes, as well as add the presence of closed string three-form fluxes in the background. Both effects will in general lift the D7-brane geometric moduli \cite{Jockers:2005zy,Gomis:2005wc,Martucci:2006ij}, while the D7-brane Wilson line moduli will remain unaffected. In contrast, the BPS conditions for D6-branes do not allow for non-trivial bundles, and considering more general compactifications to 4d Minkowski will not change the number of D6-brane moduli, which will be still be given by $b_1(\Pi_3)$ \cite{Marchesano:2006ns,Koerber:2006hh}. This has dramatic consequences for model building, where one is typically forced to either consider D6-branes wrapping three-spheres/Lens spaces or fractional D6-branes stuck at orbifold fixed points \cite{Blumenhagen:2005tn}.

While the above counting of D-brane moduli is widely accepted in the literature of type II string compactifications, the purpose of this paper is to show that it is not correct. The key observation is that the content of table \ref{t:BPS} has been obtained by considering a D-brane action in a frozen closed string background. While this gives necessary conditions for a D-brane configuration to be BPS, unfreezing the bulk degrees of freedom may unveil additional constraints that result into a potential for some of the open string moduli of table \ref{t:BPS}. For instance let us consider the following superpotential 
\be
W\, =\, \Phi_{\rm open} \cdot X_{\rm closed}
\label{supobili}
\ee
where $\Phi_{\rm open}$ is a open string modulus in the sense of table \ref{t:BPS}, while $X_{\rm closed}$ is a closed string modulus in the absence of D-branes. The F-term condition for $\Phi_{\rm open}$ will fix the vev of $X_{\rm closed}$ to a certain value, while the F-term of $X_{\rm closed}$ will do the same for $\Phi_{\rm open}$. However, if we did not consider $X_{\rm closed}$ as a dynamical field, the corresponding piece of the F-term potential would not show up in our description and so it would wrongly appear that $\Phi_{\rm open}$ does not develop a potential.

In the following sections we will show that a superpotential of the form (\ref{supobili}) is indeed generated for certain D-branes in Calabi-Yau compactifications, with $\Phi_{\rm open}$ being among those moduli in table \ref{t:BPS} that involve Wilson lines. As a consequence, some of these D6 and D7-brane moduli will be lifted even if a Dirac-Born-Infeld analysis cannot detect the generation of their potential. 

Rather than in terms of an effective superpotential, we will mainly focus on describing D-brane moduli lifting from a microscopic point of view. In this sense, we will see that the reason for moduli lifting is a combination of D-brane backreaction and flux quantisation in compact cycles of $\CM_6$. More precisely, we find that due to flux quantisation, the RR flux induced by D-brane backreaction is only compatible with supersymmetry for a subset of values of the D-brane moduli. Away from such locus the backreaction will not satisfy the bulk supersymmetry conditions and this will increase the energy of the configuration.

This effect is particularly manifest for the case of D6-brane position moduli, whose deformations modifies the backreacted RR flux $F_2$. Then, by holomorphicity, a similar effect should apply to the D6-brane Wilson lines that complexity such positions. This is somewhat more surprising, since Wilson line vevs do not appear neither in the D-brane action nor in the Bianchi identities for bulk field strengths. Hence, it is not obvious how the energy of the supergravity plus D-brane system could depend of them. Nevertheless, using again flux quantisation we argue that certain Wilson lines do indeed backreact, in the sense that changing their vev does also shift the value of RR background fields by a closed form, and this generates a potential for them. Such observation applies to both D6 and D7-brane Wilson lines and we find that in both cases the Wilson lines are lifted if a simple topological condition is met. Namely, that they are Poincar\'e dual in $\Pi_{p-3}$ to a $\pi_{p-2}$ cycle which is homologically nontrivial in $\CM_6$. 

The paper is organised as follows. In section \ref{s:IIA} we discuss the backreaction of D6-branes in special Lagrangian submanifold of a compact manifold, showing that  this backreaction is not supersymmetric for arbitrary locations. In section \ref{s:supoa} we reproduce the same effect in terms of a D6-brane effective superpotential that takes the form (\ref{supobili}). In section \ref{s:WLS} we describe the microscopic mechanism by which certain D6-brane Wilson lines can backreact and therefore be lifted. In section \ref{s:D7WL} we extend our analysis to those D7-brane Wilson lines that suffer the same fate. In section \ref{s:conclu} we present our conclusions and briefly discuss possible applications of our findings. Finally several technical details and definitions have been relegated to the appendices \ref{susy} and \ref{Leq}.

\section{D6-brane backreaction in compact spaces}
\label{s:IIA}

The backreaction of $N$ D6-branes in flat space is given by
\bes
\begin{align}
 ds^2 & \, = \, e^{2A} \eta_{\mu\nu} dx^\mu dx^\nu + e^{-2A} \d_{ij} dy^idy^j\\
C_{7}   &\, = \, g_s^{-1} (e^{4A} - 1)\, dx^0 \wedge \dots \wedge dx^6 \\
e^\phi  & \, = \, g_s e^{3A} \\
e^{-4A}  & \, = \, 1 + \frac{r_6}{r}
\end{align}
\label{backflat}
\ees
where $x^\mu$, $\mu =0, \dots 6$ are coordinates parallel to the D6-brane while $y^i$, $i = 7,8,9$ are transverse to it. In addition $r $ is the radial distance to the D6-brane locus, $g_s = e^{\phi_0}$ the asymptotic value of the string coupling, and $r_6 = \rho_6 g_s N l_6$, with $\rho_6$ a numerical factor that will not be relevant in the following. 

One can rewrite this solution in terms of the Ansatz for type IIA 4d Minkowski vacua
\be
ds^2 \, = \, e^{2A} ds^2_{\IR^{1,3}} + ds^2_{X_6}
\label{4da}
\ee
where $X_6$ is the 6d internal manifold on whose coordinates the warp factor $A$ depends. It is then easy to check that this solution satisfies the supersymmetry conditions 
\bes
\begin{align}
\label{fluxrel1}
d(3A- \phi)\, =\, H_{\rm NS} + i dJ\, =\,  0 & \quad & F_0\, =\, \tilde F_4\, =\, \tilde F_6 \, =\, 0  \\
d(e^{2A-\phi}\im\, \Omega)\, =\, 0 & \quad & d(e^{4A-\phi} \re\, \Omega)\, =\,  - e^{4A} *_6 F_2
\label{fluxrel2}
\end{align}
\label{fluxrel}
\ees
where $*_6$ is the Hodge star operator in the internal space $X_6$, including the warp factor. The fact that this supergravity background is supersymmetric is an indication that the backreacted D6-brane is BPS and preserves some of the supersymmetry of the initial background (in this case 10d Minkowski flat space). We may now add further D6-branes that also contain the coordinates of $\IR^{1,3}$, and which are either parallel to the initial one or intersect it at supersymmetric SU(3) angles in the remaining coordinates \cite{bdl96}. By backreacting them we will obtain a more complicated supergravity solution which will nevertheless satisfy (\ref{4da}) and (\ref{fluxrel}). Again, this indicates that this system of intersecting D6-branes is mutually BPS and at least $\CN=1$ 4d supersymmetry is preserved by it. Finally, the fact that we can displace transversely the D6-brane locations without spoiling (\ref{fluxrel}) corresponds to the fact that there is no force between these mutually BPS D-branes, and that there is a set of flat directions that can interpreted as open string moduli. 

In principle, one may expect a similar picture to apply if instead of $\IR^{1,9}$ we consider $\IR^{1,3} \times \CM_6$, where $\CM_6$ is a a non-compact Calabi-Yau manifold. Indeed, if we backreact a D6-brane along $\IR^{1,3}$ and a submanifold $\Pi_3 \subset \CM_6$ we will be sourcing a RR two-form flux which locally satisfies 
\be
dF_2\, =\, \d_3(\Pi_3)
\ee
where $\d_3$ is delta-function three-form with support on $\Pi_3$ and indices transverse to it. Similarly the D6-brane backreaction will source the dilaton, metric and warp factor in such a way that we have an Ansatz of the form (\ref{4da}), where $X_6$ is identical to $\CM_6$ in terms of differentiable manifolds but endowed with the backreacted metric. Finally, if $\Pi_3$ is a special Lagrangian calibrated by $\re\, \Omega$, then eqs.(\ref{fluxrel}) will be satisfied.

On the other hand, new restrictions may arise when $\CM_6$ is a compact manifold with non-trivial topology. First of all, compactness of $\CM_6$ implies that the total D6-brane charge needs to cancel, simply because $F_2$ is globally well-defined. More precisely we have a Bianchi identity of the form
\be
dF_2\, =\, \sum_\a \d(\Pi_3^\a) + \d(\Pi_3^{\a^*}) - 4 \d(\Pi_3^{O6})
\label{backF2}
\ee
where the index $\a$ runs over the D6-branes of a given compactification, $\a^*$ over their orientifold images and $\Pi_3^{O6}$ stand for the O6-plane loci. This equation will have a solution for a globally well-defined $F_2$ if and only if the following  equation in $H_3(\CM_6)$ is satisfied 
\be
\sum_\a [\Pi_3^\a] + [\Pi_3^{\a^*}] - 4 [\Pi_3^{O6}]\, =\, 0
\label{RRD6}
\ee
which is usually known as the RR tadpole condition \cite{Aldazabal:2000dg}. In addition, for the wavefunction of a D0-brane to be well-defined, the integral of $F_2$ over any two-cycle must be quantised. More precisely, over each two-cycle $\pi_2^a \subset \CM_6-\Pi_3^{D6}$ (with $\Pi_3^{D6}$ is the sum of all the three-cycles wrapped by D6-branes and O6-planes) we must have
\be
n_a\,=\, \frac{1}{\l_s} \int_{\pi_2^a} F_2\, \in \, \IZ
\label{quantF2}
\ee
where $\l_s = 2\pi \sqrt{\a'}$ is the string length. This condition not only applies to those two-cycles that surround a D6-brane but also to the non-trivial two-cycles of $\CM_6$ that also belong to $\CM_6 - \Pi_3^{D6}$. As we will see, it is due to imposing (\ref{quantF2}) to the latter two-cycles where new restrictions in moduli space appear. 

These consistency conditions are of topological nature, but supersymmetry imposes further constraints on $F_2$. Indeed, notice that one can rewrite the second equation in (\ref{fluxrel2}) as
\be
F_2\, =\, *_{10} d *_{10} \left( e^{-\phi} \im\, \Omega\right)
\label{coexact}
\ee
which means that $F_2$, seen as a two-form in the full 10d backreacted space, is co-exact since $e^{-\phi} \im\, \Omega$ is globally well-defined. In general, similarly to Hodge decomposition one can split any two-form with legs in the internal six-dimensional space as
\be
F_2\, =\, d\a_1 + F_2^{\rm harm} + *_{10} d *_{10} \g_3
\label{Hdecomp}
\ee
that is into an exact, harmonic and co-exact pieces. Notice that the case of $F_2$ the three-form $\g_3$ is fixed by the Bianchi identity (\ref{backF2}) while the other two components are not. The additional input of supersymmetry is then that $\g_3 = e^{-\phi} \im\, \Omega$ and $\a_1 = F_2^{\rm harm} =0$.

On the one hand, that $\a_1$ vanishes is easy to achieve, as one can always adjust the background value of the RR potential $C_1$ in order to cancel such component.  On the other hand, requiring that $F_2^{\rm harm}$ vanishes is non-trivial, due to the quantisation condition (\ref{quantF2}). Indeed, because in general the integral of $d^*\g_3$ will be non-vanishing and non-integer over the non-trivial two-cycles of $\CM_6$, one needs to include a harmonic piece in $F_2$ such that the full integral adds up to an integer, as required by consistency. If that is the case, the D6-brane configuration will be non-supersymmetric even if all D6-branes wrap special Lagrangian three-cycles. As we will discuss below the backreacted $F_2^{\rm harm}$ will depend on certain D6-brane locations which means that, at the end of the day, supersymmetry will impose a constraint in the D6-brane moduli space. One then expects that the number of constraints imposed by supersymmetry can be up to $b_2 (\CM_6) = {\rm dim} \,H_2(\CM_6, \IR)$, which measures the number of independent two-cycles in $\CM_6$.\footnote{In fact, in orientifold compactifications $F_2$ is a two-form odd under the geometric orientifold action, so the maximal number of constraints is actually given by $b_2^- (\CM_6) = {\rm dim}\, H_2^-(\CM_6, \IR)$.} 

Rather than computing the value of $F_2^{\rm harm}$ for a specific compactification with backreacted D6-branes, let us discuss how does it depend on the D6-brane locations. More precisely, we will show that if (\ref{coexact}) is satisfied, it cannot be so after changing the location of certain D6-branes. It turns out that the expression (\ref{coexact}) is not the most suitable one for such analysis, simply because it depends on $*_{10}$, which in turn depends on the warp factor and ultimately on the D6-brane locations. Instead,\footnote{We would like to thank L.~Martucci for discussions regarding this point.} one can use the equivalent condition formulated in terms of a generalised Dolbeault operator $d^{\CJ}$, namely \cite{Tomasiello:2007zq}
\be
F_2\, =\, d^{\CJ} (e^{-\phi} \re\, \Omega) 
\label{coJexact})
\ee
where the definition of $d^{\CJ}$ is given in Appendix \ref{susy}. For our purposes here it suffices to point out that for the type IIA backgrounds at hand this condition can be rewritten as \cite{Martucci:2009sf}
\be
d(e^{-\phi} \re\, \Omega)\, =\, - J \wedge F_2
\label{fluxrelJ}
\ee
which substitutes the second equation in (\ref{fluxrel2}). The important point is that the two-form $J$ does not depend on the D6-brane backreaction or their location, and in fact it remains the same as in the unbackreacted space $\CM_6$. 

Hence, in order to see if the D6-brane backreaction satisfies the supersymmetry equation (\ref{fluxrelJ}), we may consider the Calabi-Yau orientifold $\CM_6$ with D6-branes wrapping special Lagrangian three-cycles $\Pi_3^a$ in it and a quantised two-form flux satisfying (\ref{backF2}). Let us for instance take a D6-brane wrapping the special Lagrangian $\Pi_3$ and displace it to the new special Lagrangian three-cycle $\Pi_3'$ homotopic to the former. The new backreacted flux $F_2'$ will be given by
\be
F_2' \, =\,  F_2 + \Delta F_2
\ee
where $F_2$ and $F_2'$ solve for eq.(\ref{backF2}) before and after moving the D6-brane, respectively. Both $F_2$ and $F_2'$ are quantised two-forms, so $\Delta F_2$ is quantised as well, and it satisfies the equation
\be
d \Delta F_2\, =\, \d (\Pi_3') - \d (\Pi_3)
\label{DF2}
\ee

We would now like to check whether for some particular D6-brane displacement we have that
\be
\int_{\CM_6} \Delta (F_2 \wedge J) \wedge \omega_2 \, =\, \int_{\CM_6} \Delta F_2 \wedge J \wedge \omega_2\, \neq \, 0 
\label{int6}
\ee
for some closed two-form $\omega_2$ of $\CM_6$, where we have used that $J$ does not depend on the D6-brane location. If the rhs of (\ref{int6}) does not vanish for some closed $\omega_2$ it means that $J \wedge F_2$ cannot be written as an exact form either before or after displacing the D6-brane, and that supersymmetry is broken for D6-brane deformations of this sort.

To proceed, one may follow \cite{Marchesano:2014bia} (see also \cite{Hitchin99}) and use that $F_2$ is quantised, it satisfies eq.(\ref{backF2}) and that $J\wedge \omega_2$ is closed to derive the identity
\be
\int_{\CM_6} \Delta F_2 \wedge J\wedge \omega_2 \, =\, \int_{\Sigma_4} J \wedge \omega_2
\label{int4chain}
\ee
where $\Sigma_4$ is a four-chain connecting $\Pi_3$ and $\Pi_3'$. Notice that if $\omega_2$ is exact and these three-cycles are Lagrangian the integrals will identically vanish, so the only way to obtain a non-vanishing result is if $\omega_2$ contains a harmonic two-form of $\CM_6$. In this sense, the integrals (\ref{int4chain}) measure how the harmonic component of $\Delta F_2$ in $\CM_6$ changes with the D6-brane location. In the language of \cite{Hitchin99,Marchesano:2014bia}, we see that the integral vanishes and $\Delta F_2$ has no harmonic piece only if the three-cycles $\Pi_3^a$ and $\Pi_3^{a\, '}$ are linearly equivalent.

To gain further insight into the condition (\ref{int6}) let us take $\Pi_3'$ to be the infinitesimal deformation of $\Pi_3$ by a normal vector $X$. Using again that $J|_{\Pi_3} = 0$ we have that the chain integral becomes
\be
\int_{\Pi_3} \iota_XJ \wedge \omega_2
\label{iotaJ}
\ee
By McLean's theorem \cite{McLean} we know that in order for $X$ to describe a special Lagrangian deformation, $\iota_XJ$ must be a harmonic one-form in $\Pi_3$ which we can take to be integral. We can then use Poincar\'e duality to write the above expression as
\be
\int_{\pi_2^X} \omega_2
\label{findef}
\ee
where $\pi_2^X$ is a two-cycle of $\Pi_3$ in the Poincar\'e dual class of $\iota_XJ$. We then find that an infinitesimal deformation $X$ of $\Pi_3$ violates the supersymmetry condition (\ref{fluxrelJ}) if and only if the integral (\ref{findef}) does not vanish. This implies in particular that $\Pi_3$ must have a two-cycle $\pi_2$ which is non-trivial in the ambient space $\CM_6$. Then, deforming the D6-brane location along the direction that corresponds to such two-cycle will break supersymmetry because it modifies the harmonic piece of the backreacted $F_2$ in the Calabi-Yau metric $\CM_6$. Finally, it is easy to check that $F_2$ is odd under the orientifold involution defined on $\CM_6$, and so must be $\omega_2$ and $\pi_2^X$ in order for the above integrals not to vanish. 

Hence, the final picture is that by deforming the location of D6-branes that contain non-trivial odd two-cycles in the compactification space  $\CM_6$ a harmonic piece will be generated for a quantised flux $F_2$ satisfying (\ref{quantF2}) and supersymmetry will be broken. From the viewpoint of the fully backreacted supergravity background we will have that the internal metric, warp factor and dilaton will be sourced such that 
\be
d [e^{-4A} *_6 d(e^{4A-\phi}\re \Omega)]\, =\, - \sum_\a \d(\Pi_3^\a) + \d(\Pi_3^{\a^*}) - 4 \d(\Pi_3^{O6})
\ee
The backreacted flux $F_2$ will satisfy a similar Poisson equation, but due to quantisation it will also contain a harmonic piece $F_2^{\rm harm}$ in the 10d decomposition (\ref{Hdecomp}) which prevents eq.(\ref{coexact}) to be satisfied. Such component will raise the energy of the compactification via the 4d effective potential computed in \cite{Lust:2008zd} 
\be
\CV_{\rm eff}\, =\, \oh \int_{X_6} d{\rm vol}_6 e^{4A} \left[*_6 F_2 + e^{-4A} d (e^{4A-\phi}   \re \Omega) \right]^2 +\dots
\ee
where we the remaining pieces of the potential are a sum of squares not relevant for the present discussion. More precisely one finds that
\be
\CV_{\rm eff}\, =\, \oh \int_{X_6} d{\rm vol}_6\, e^{4A} F_2^{\rm harm} \wedge *_{6} F_2^{\rm harm}
\label{veffh}
\ee
where $F_2^{\rm harm}$ depends on the D6-brane position as described above. Hence, such would-be open string moduli pick up a mass, even if they preserve the D6-brane special Lagrangian condition. In particular, they will be fixed to those values such that $F_2^{\rm harm}$ vanishes.

\section{Superpotential analysis}
\label{s:supoa}

Let us now show that, from a 4d macroscopic viewpoint, the effective potential (\ref{veffh}) arises from a superpotential bilinear in open and closed string fields. This can be done by considering the classical D6-brane superpotential\footnote{By classical we mean the superpotential that arises before taking into account worldsheet instantons, most precisely holomorphic disk instantons ending on the D6-brane one-cycles.}
%Such superpotential will be a good approximation whenever such one-cycles are large enough.} 
\cite{Thomas:2001ve,Martucci:2006ij}
\be
\Delta W_{\rm clas}^{D6}\, =\, \int_{\Sigma_4} (J_c + F)^2
\label{supoluca}
\ee
where $\Sigma_4$ is a four-chain connecting the three-cycle $\Pi_3$ and a homotopic deformation $\Pi_3'$. If the deformation is infinitesimal and given by the normal vector $X$ we can write $W$ as \cite{Thomas:2001ve}
\be
\Delta W_{\rm clas}^{D6}\, =\, \int_{\Pi_3} (J_c + F) \wedge (\iota_X J_c + A)
\label{supoD6}
\ee
We may now expand the complexified K\"ahler form as
\be
J_c\, =\, B+iJ \, =\,  T_a\, \omega_2^a
\ee
where $\{ \omega_2^a\}$ is a basis of integer harmonic two-forms of $\CM_6$, and $T_a$ are the corresponding  K\"ahler moduli. In addition we may define the open string deformation as 
\be
\Phi_{\rm D6}\, =\, \iota_X J_c|_{\Pi_3} + A \, =\, (\th^j+\lam_i^j\phi^i) \zeta_j \, =\, \Phi_{\rm D6}^j \zeta_j
\ee
where $\zeta_j/2\pi$ is a quantised harmonic one-form of $\Pi_3$ and (see e.g. \cite{kt11,Marchesano:2014bia} for details)
\be
X\, =\, \phi^j X_j \quad \quad A\, =\, \frac{\pi}{l_s} \theta^j \zeta_j \quad \quad \iota_{X_i}J_c|_{\Pi_3}=\lam_i^j\zeta_j
\ee
Because $\zeta_j$ is harmonic, the field $\Phi^j_{\rm D6}$ corresponds to a D6-brane deformation that preserves the worldvolume supersymmetry conditions 
\be
J_c|_{\Pi_3} + F\, =\, 0 \quad \quad \im\, \Omega|_{\Pi_3}\, =\, 0
\label{SUSYD6}
\ee
and so it is typically identified with an open string modulus. However, plugging both expressions into (\ref{supoD6}) we obtain the non-trivial superpotential
\be
\Delta W_{\rm clas}^{D6}\, =\, m_j^a\, \Phi_{\rm D6}^j T_a \quad \quad {\rm with} \quad \quad m_j^a\, =\, \int_{\Pi_3} \omega_2^a \wedge \zeta_j
\label{supobil}
\ee
which, as announced, is a bilinear on the open string $\Phi_{D6}^j$ and closed string $T_a$ fields. Notice that the superpotential is only non-trivial if the integer numbers $m_j^a$ are non-vanishing. Recalling the discussion around eq.(\ref{findef}), it is easy to see that $m_j^a\neq 0$ if and only if some two-cycle of $\Pi_3$ is also non-trivial in the ambient space as an element of $H_2(\CM_6, \IR)$. Precisely when this happens, some of these naive moduli will be stabilised by an F-term scalar potential, in agreement with the results obtained in the previous section.

In order to see this in some detail let us consider the simple case where $\Pi_3$ has just one non-trivial harmonic form $\zeta$ and its dual two-cycle $\pi_2$ is such that $\int_{\pi_2} \omega_2^a = - \int_{\pi_2} \omega_2^b = 1$, while all the other bulk two-forms integrate to zero. Then we have that the superpotential reads
\be
\Delta W_{\rm clas}^{D6}\, =\, \Phi_{D6} (T_a - T_b)
\label{supoex}
\ee
whose critical points are given by $T_a = T_b$ and $\Phi_{D6} = 0$. Hence, we recover that one open string modulus and one linear combination of closed string moduli are fixed by this superpotential. 

Care should be taken when deriving a scalar potential from (\ref{supoex}) since, as pointed out in \cite{Martucci:2006ij}, the expression (\ref{supoluca}) is only the difference of the superpotential between two D6-brane positions, and not the absolute $W^{\rm D6}_{\rm clas}$. Nevertheless, let us assume that the full superpotential is such that we have a definite-positive, no-scale potential as in \cite{Giddings:2001yu} and that the D6-brane configuration is such that $W_{\rm clas} = 0$ and we are at a supersymmetric minimum. Then the piece of scalar potential that we obtain from (\ref{supoex}) reads
\be
\CV\, =\, e^{K}\left( K^{\Phi\Phi} |T|^2 + K^{TT} |\Phi_{D6}|^2 \right)
\ee
where $T \equiv T_a -T_b$. Hence, we find that the closed and the open string modulus are fixed to the values $T_a = T_b$ and $\Phi_{D6}=0$ separately. 

On the one hand, the potential for the closed string modulus $T$ is easy to interpret. Indeed, whenever $T_a \neq T_b$ the pull-back of $J_c$ on $\Pi_3$ will be non-vanishing, and so the supersymmetry conditions (\ref{SUSYD6}) will not be met. Moreover, the integral of $\omega_2^a$ and $\omega_2^b$ over $\pi_2 \subset \Pi_3$ will not change if we deform the D6-brane embedding. Hence, a D6-brane wrapping $\Pi_3$ will break supersymmetry and have an excess of energy unless $T_a = T_b$. In fact, the piece of the potential that goes like $|T|^2$ can be easily derived from the analysis of D6-brane DBI action in such background, following similar steps as those performed in section 5.2 of \cite{Font:2006na} for coisotropic D8-branes. 

On the other hand, the potential for the open string modulus $\Phi_{D6}$ cannot be derived from a DBI analysis. Indeed, since such term arises form the F-term of the closed string modulus $T$, it will not appear if $T$ is not considered dynamical. But not considering $T$ as dynamical is precisely what is done when we analyse a D6-brane action in a frozen closed string background. Hence, it is only via D6-brane backreaction effects of the bulk that we can understand the nature of this piece of the potential, as done in the previous section. 

An interesting byproduct of last section analysis is that it allows to deduce the D6-brane classical superpotential in absolute terms, instead of defining just $\Delta W_{\rm clas}^{\rm D6}$. Indeed, recall that the crucial supersymmetry condition for the above analysis can be rewritten as (\ref{fluxrelJ}), which implies that $J \wedge F_2$ is exact in the cohomology of $\CM_6$. In fact, as we will discuss in the next section, one can use the remaining supersymmetry conditions to argue that $J_c \wedge F_2$ must be exact as well. This is satisfied if and only if
\be
\int_{\CM_6} F_2 \wedge J_c \wedge \omega_2^a = 0 \quad \quad \forall \, [\omega_2^a] \in H^2(\CM_6,\IR)
\ee
In terms of a superpotential, this condition can be derived by replacing $ \omega_2^a \raw J_c = T_a  \omega_2^a$ and imposing the F-term condition on each K\"ahler modulus $T_a$ separately. Hence, we are led to the expression
\be
W_{\rm clas}^{\rm D6}\,=\, \int_{\CM_6} F_2 \wedge J_c \wedge J_c \, =\,  \int_{\Sigma_4^{\rm tot}} J_c^2
\ee
which is quite familiar from the type IIA literature. Indeed, the first expression for $W_{\rm clas}^{\rm D6}$ is nothing but the standard flux superpotential for type IIA Minkowski vacua \cite{superIIA}, where now $F_2 = F_2^{\rm D6}$ stands for the flux coming from the backreaction of D6-branes and O6-planes. Of course one can also add to this backreacted flux a quantised background flux $F_2^{\rm bkg}$. The sum of both fluxes will enter the superpotential as $F_2 = F_2^{\rm D6} + F_2^{\rm bkg}$, and we can understand both contributions as the sum $W_{\rm brane} + W_{\rm flux}$ discussed in e.g. \cite{Lerche:2003hs}.

The second expression for $W_{\rm clas}^{\rm D6}$ is obtained by replacing the current $F_2^{\rm D6}$ by a dual four-chain $\Sigma_4^{\rm tot}$ which connects all D6-branes and O6-planes. That such four-chain exists is a direct consequence of the RR tadpole condition (\ref{RRD6}), and by focusing on a single D6-brane one obtains the expression (\ref{supoluca}) on which the analysis os this section is based.

In fact, expression (\ref{supoluca}) is obtained after replacing $J_c \raw J_c + F$, where $F=dA$ will contain the Wilson line dependence of the superpotential. In the next section we will discuss how this replacement should arise. In any case, from this superpotential analysis it is clear that certain D6-brane Wilson lines should also be stabilised, since they enter $\Phi_{D6}$ as the complexification of the position moduli which get affected by the superpotential (\ref{supobil}). This may sound surprising, since typically it is assumed that D-brane Wilson lines are free of any scalar potential, and that the D-brane and background configuration is fully independent of them. In the following we will show that this intuition is wrong, and that there is a quite simple microscopic mechanism by which Wilson lines are stabilised. 

\section{Wilson line moduli stabilisation}
\label{s:WLS}

An important observation is that the supersymmetry condition (\ref{fluxrelJ}) may not be the only that is spoiled when changing the location of a D6-brane over its naive moduli space. Indeed, a different supersymmetry condition that also turns out to be relevant is
\be
\tilde F_4\, =\, dC_3 - C_1 \wedge H_{NS} \, =\, 0
\label{tF4}
\ee
In general the gauge invariant flux $\tilde F_4$ satisfies the Bianchi identity
\be
d\tilde F_4 + F_2 \wedge H_{NS} \, =\, j_{\rm D4}
\ee
with $j_{\rm D4}$ a five-form current describing the D4-brane charge carried by the D-branes of the configuration. In the case of compactifications with O6-planes, D-brane BPSness forbids the presence of D4-branes, while D6-branes must carry a vanishing worldvolume flux $\CF = B + F$. Hence there is no induced D4-brane charge and so $j_{\rm D4} = 0$.\footnote{Coisotropic D8-branes will in general violate this condition, since they do carry induced D4-brane charge \cite{Font:2006na}. For simplicity we will not consider their presence here.} In addition, we have that  an independent supersymmetry condition imposes that $H_{NS} = 0$. As a result, for each point of the naive moduli space of special Lagrangian D6-branes we have that the backreacted background satisfies
\be
d\tilde F_4\, =\, 0
\label{tF4c}
\ee
which of course also applies for any choice of Wilson lines on such D6-branes. In general, Wilson lines do not enter into the Bianchi identity of any background flux, which typically leads to the intuition that they do not backreact into the closed string background. In the following we will argue that such intuition is wrong by considering the quantisation conditions that $\tilde F_4$ must satisfy in compact manifolds. 

As discussed in \cite{Marolf:2000cb} the gauge invariant four-form flux $\tilde F_4$ does not satisfy a quantisation condition itself, but we must instead consider the notion of Page charge and take the combination $\tilde F_4 + F_2 \wedge B$. Then we have that the proper quantisation condition reads
\be
\frac{1}{\l_s^3} \int_{\pi_4} \tilde{F}_4 + F_2 \wedge B\, \in\, \IZ 
\label{quantF4}
\ee
over each four-cycle $\pi_4 \subset \CM_6-\Pi_3^{D6}$. Here $F_2$ is the previous two-form RR flux that arises from the backreaction of D6-branes. As such we have that $d(F_2 \wedge B) = dF_2 \wedge B = 0$ since due to eq.(\ref{SUSYD6}) the pull-back of the B-field vanishes on each D6-brane. Hence $\tilde F_4 + F_2 \wedge B$ is a closed, quantised four-form whose integral over each four-cycle does not change when we move on the naive D6-brane moduli space. 

While the sum $\tilde F_4 + F_2 \wedge B$ is quantised, both factors may not be so separately. If $F_2 \wedge B$ is not quantised it means that it contains a non-integer harmonic four-form piece, and so $\tilde F_4$ must also contain a non-trivial harmonic four-form in order to satisfy (\ref{quantF4}). This means in particular that $\tilde F_4 \neq 0$ and so supersymmetry is broken. 

Following the same philosophy of section \ref{s:IIA}, let us consider the case where $F_2 \wedge B$ is a quantised four-form\footnote{Using large gauge transformations of the B-field, one may simply consider the case where $F_2 \wedge B$ is exact, which is the supersymmetry condition used in the previous section.} and let us see whether moving in the naive D6-brane moduli space spoils this condition. As before we consider the case in which we change a D6-brane location from $\Pi_3$ to $\Pi_3'$, and define the corresponding difference of two-form flux $\Delta F_2$ satisfying (\ref{DF2}). This does not change the B-field at all, and so there is a change in the harmonic component of $F_2 \wedge B$ if
\be
\int_{\CM_6} \Delta F_2 \wedge B \wedge \omega_2 \, =\, \int_{\Sigma_4} B \wedge \omega_2\, \neq\, 0
\label{int4chainB}
\ee
where $\omega_2$ is some harmonic two-form of $\CM_6$, and $\Sigma_4$ has been defined as in (\ref{int4chain}). Taking now an infinitesimal deformation given by the normal vector $X$, the change in $F_2 \wedge B$ will be measured by 
\be
\int_{\Pi_3} \iota_XB \wedge \omega_2
\label{iotaB}
\ee
which is similar to eq.(\ref{iotaB}) with the replacement $J \raw B$. Again, this change will be non-vanishing whenever the three-cycle $\Pi_3$ contains a two-cycle $\pi_2^X$ that is non-trivial in the odd homology of $\CM_6$. 

So far we have only proven that when deforming certain D6-branes along their special Lagrangian moduli space one can break supersymmetry in two independent ways, by switching on a non-exact component for $J \wedge F_2$ and a non-exact component for $B \wedge F_2$ or equivalently for $\tilde F_4$. By holomorphicity in the open string modulus $\Phi_{\rm D6} = \iota_X J_c +A$, we would expect in these cases there is also a potential for the D6-brane Wilson line moduli. 

Indeed, instead of changing the location of the D6-brane three-cycle $\Pi_3$ let us perform a change in its Wilson line $\Delta A = A' - A$. By a gauge transformation we can transform such change in a shift of the B-field by a exact form $\Delta B = B' - B$ such that
\be
\Delta B\, =\, d \Theta_1 \quad \quad {\rm and} \quad \quad \Theta_1|_{\Pi_3}\, =\, \Delta A 
\ee
Such shift in the B-field by an exact form will not increase the energy of the system via the NS-flux potential $\int |H_{NS}|^2$. However, since $F_2$ is not closed it may change the harmonic piece of $B \wedge F_2$ and so increase the energy by shifting $\tilde F_4$. Indeed, we find that such change is measured by
\be
\int_{\CM_6} \Delta (F_2 \wedge B) \wedge \omega_2\, =\, \int_{\CM_6} F_2 \wedge \Delta B \wedge \omega_2\, =\, - \int_{\CM_6} dF_2 \wedge \Delta A \wedge \omega_2\, =\,  \int_{\Pi_3} \Delta A \wedge \omega_2
\ee
and so it will not vanish whenever $\Delta A$ is Poincar\'e dual to a two-cycle of $\Pi_3$ non-trivial in $H_2^-(\CM_6,\IR)$, in agreement with our previous results.

To summarise, we have found that certain D6-brane Wilson lines also `backreact' into a harmonic component for $\tilde F_4$, which increases the energy of the system and fixes their value. Notice that this shift of $\tilde F_4$ is compatible with all the 10d Bianchi identities and equations of motion. In fact, one can argue that the background value of $\tilde F_4$ needs to change as described by looking at the domain wall solution interpolating between different D6-brane configurations. 

Indeed, let us first consider the case where the D6-brane location is changed from $\Pi_3$ to $\Pi_3'$. The domain wall connecting these two configurations will be a D6-brane wrapped on the chain $\Sigma_4$ connecting both three-cycles and localised in the 4d coordinate $x^3$. Now, if (\ref{int4chainB}) is true it means that the domain wall D6-brane will be magnetised by the presence of the B-field, and so it will actually be a D6/D4-brane bound state. As such, not only the background value of $F_2$ will change when we cross this domain wall, but also that of $\tilde F_4$. Finally, adding a relative Wilson line between $\Pi_3$ and $\Pi_3'$ will result in a worldvolume flux $F$ threading the D6-brane domain wall, whose total D4-brane charge will be induced by $\CF = B+F$. One then obtains again that the D6-brane Wilson lines shift the value of $\tilde F_4$, although only if via Poincar\'e duality on $\Pi_3$ they correspond to a non-trivial odd two-cycle of $\CM_6$. 

It is easy to see this mechanism for stabilising Wilson lines is more general than the type IIA setup that we are discussing, and that it can in principle be applied to other kind of string vacua as well. In the following we will briefly comment on how it allows to stabilise Wilson lines in type IIB vacua with D7-branes. 

\section{Stabilising D7-brane Wilson lines}
\label{s:D7WL}

Let us consider a type IIB orientifold compactification with O3/O7-planes, and with space-time filling D7-branes wrapping divisors of a Calabi-Yau threefold $\CW_6$. Let us in particular consider a D7-brane wrapping a divisor $S_4 \subset \CW_6$ such that $S_4$ contain harmonic one-forms or, in other words, that it contains Wilson line moduli. By Poincar\'e duality $S_4$ will contain non-trivial three-cycles and, since it is a complex submanifold, the number of independent odd-cycles must be even. So the minimal setup that we may consider is that $S_4$ contains two three-cycles $\{\pi_3^1,\pi_3^2\}$.

For our discussion, the key point is whether the three-cycles $\{\pi_3^1,\pi_3^2\}$ are trivial in $\CW_6$ or not. If they are non-trivial then one can follow a discussion parallel to the one carried for the D6-brane, and argue that such D7-brane develops a superpotential of the form
\be
W\, =\, \Phi_{\rm D7} \cdot X_{\rm closed}
\label{supooc}
\ee
where $\Phi_{\rm D7}$ is the complexified Wilson line that corresponds to these three-cycles, and $X_{\rm closed}$ is a linear combination of complex structure moduli of $\CW_6$. 

As before, the obstruction to move on closed string moduli space that arise from (\ref{supooc}) can be derived by analysing the D-brane supersymmetry conditions, which in this case impose that
\be
\Omega|_{S_4} \, =\, 0
\label{D7cpx}
\ee
which is equivalent to ask that $S_4$ is holomorphic. The three-form $\Omega$ can be understood as a linear combination of integer three-forms whose coefficients depend on the complex structure moduli of $\CW_6$. More precisely we have that the complex structure moduli are (redundantly) defined as
\be
z^A\, =\, \int_{\g_3^A} \Omega \quad \quad \omega^B\, =\,  \int_{\g_3^B} \Omega
\ee
where $\{\g_3^A, \g_3^B\}$ is a symplectic basis of integer three-cycles of $\CW_6$ with $[\g_3^A] \cdot [\g_3^B] \, =\, \d^{AB}$.

As we move along the complex structure moduli space the integral of $\Omega$ over the non-trivial three-cycles of $\CW_6$ changes. Hence, if $S_4$ contains any of these non-trivial three-cycles we may reach a point in which 
\be
\int_{\pi_3} \Omega \neq 0
\ee
so that the four-cycle $S_4$ is no longer holomorphic. Hence, as already pointed out in \cite{Jockers:2004yj}, the corresponding complex structure deformation should be obstructed.

The Wilson line obstruction can be seen from considering the superpotential
\be
W^{\rm D7}\, =\, \int_{\Sigma_5} \Omega \wedge F
\label{supoD7}
\ee
which is the analogue of the D6-brane superpotential of section \ref{s:supoa}. Taking $\Sigma_5$ the five-chain that connects all D7-branes and O7-planes and using Stokes' theorem we obtain that
\be
D_\a W^{\rm D7} \,=\,  \sum_i \int_{S_4^i} \chi_\a \wedge A^i
\ee
where $\chi_{\a}$ is a harmonic (2,1)-form of $\CW_6$ that represents a complex structure modulus \cite{Candelas:1990pi,Giddings:2001yu}, and $D_\a$ is the corresponding supergravity covariant derivative. Finally, $A^i$ is a (0,1)-form on $S_4^i$ that represents its Wilson line modulus and which will be stabilised by the scalar potential term $K^{\a\a} |D_\a W^{\rm D7} |^2$. 

Hence, as advanced, we have a superpotential of the form (\ref{supooc}) with $X_{\rm closed}$ a complex structure moduli and $\Phi_{\rm open}$ Wilson line moduli. Just like for D6-branes, this superpotential will be non-trivial only if a topological condition is met, namely that these three-cycles of $S_4$ are non-trivial also in $\CW_6$.

Finally, one can easily extend to this case the microscopic mechanism by which Wilson lines are stabilised. For type IIB flux compactifications we will have that gauge invariant three-form flux
\be
\tilde F_3 \, =\, dC_2 -C_0 H_{\rm NS}
\ee
is not quantised while the combination of three-forms $\tilde F_3 - F_1 \wedge B$ is. Switching on a Wilson line will be equivalent to shift the B-field by the appropriate exact two-form, which will nevertheless contribute to the harmonic piece of $F_1 \wedge B$ when $F_1$ is non-closed and $S_4$ contains a non-trivial three-cycle. Hence switching on such Wilson line will result on a shift of the harmonic piece of $\tilde F_3$, and this will contribute to the energy of the system via the usual scalar potential induced by background fluxes. 

\section{Conclusions and Outlook}
\label{s:conclu}

In this paper we have revisited the usual counting of D-brane moduli in type II orientifold compactifications. We have argued that such counting is inaccurate, in the sense that it may consider as moduli certain D-brane scalar fields which are fixed by compactifications effects. We have found particular examples of such fixed moduli, corresponding to D6 and D7-brane deformations that involve Wilson lines.\footnote{By Wilson line we mean the standard definition of Wilson line used in the literature, corresponding to the presence of a harmonic one-form, and not the concept of massive Wilson line formulated in \cite{msu}.} This result is particularly interesting for model building, as these D-brane scalars were thought to remain unfixed for general 4d $\CN=1$ flux compactifications. 

For instance, our analysis has shown that for a type IIA orientifold compactification with D6-branes the configuration can be non-supersymmetric even if all the D6-brane wrap special Lagrangian three-cycles. While this partially contradicts previous results in the literature (see e.g. \cite{kt07}), it is a direct consequence of imposing RR flux quantisation over the compact cycles of internal manifold $\CM_6$. In mathematical terms, one should not only require that the D6-branes and O6-planes wrap special Lagrangians, but also that this set of three-cycles is trivial in terms of linear equivalence, or rather in terms of the more general notion of D-brane linear equivalence developed in \cite{Marchesano:2014bia}, which also depends on Wilson lines. In more practical terms, we have found that a D6-brane wrapping a three-cycle $\Pi_3$ will have stabilised moduli if it contains two-cycles that are homologically non-trivial in the ambient space $\CM_6$. Finally, we have shown that such moduli fixing is reproduced by a superpotential of the form (\ref{supobili}), which can be derived from the standard expression for D6-brane superpotentials in the literature. 

While most of our analysis has been devoted to D6-branes, our reasoning also applies in other instances, as we have shown for the case of D7-brane Wilson lines. We find that such Wilson lines will be stabilised if they are Poincar\'e dual to three-cycles of $S_4$ that are non-trivial in the ambient space $\CW_6$. Again, this dynamical fixing is described by a superpotential of the form (\ref{supobili}) bilinear in open and closed string fields. This opens the possibility that these D7-branes and the above D6-branes might be related by mirror symmetry, a question to which we hope to return in the future. 

The results of this paper may be generalised in a number of ways. For instance, one may wonder which other D-branes (or NS5-branes) may develop a superpotential of the form (\ref{supobili}). In particular one may compare our results with the open-closed superpotentials involving D5-branes \cite{Lerche:2003hs,5branes}.  Also, it would be interesting to see how the D7-brane Wilson line stabilisation of section \ref{s:D7WL} is generalised to $(p,q)$-7-branes in F-theory compactifications. 

Finally, it would be interesting to see how our results can improve the construction of semi-realistic string vacua. On the one hand they should relax the model building constraints on the number of adjoint fields in D-brane models of particle physics. On the other hand, the scalar potential that arises from (\ref{supobili}) can give new possibilities for constructing string models of inflation. In particular, it allows to build new models of axion monodromy inflation \cite{future}, by sharpening the proposal made in Appendix A of \cite{msu}.

\bigskip

\bigskip

\centerline{\bf \large Acknowledgments}

\bigskip

We would like to thank Luis~E.~Ib\'a\~nez, Angel~M.~Uranga and specially Luca~Martucci for useful discussions. 
This work has been partially supported by the grant FPA2012-32828 from the MINECO, the REA grant agreement PCIG10-GA-2011-304023 from the People Programme of FP7 (Marie Curie Action), the ERC Advanced Grant SPLE under contract ERC-2012-ADG-20120216-320421 and the grant SEV-2012-0249 of the ``Centro de Excelencia Severo Ochoa" Programme. F.M. is supported by the Ram\'on y Cajal programme through the grant RYC-2009-05096. D.R. is supported through the FPU grant AP2010-5687. G.Z. is supported through a grant from ``Campus Excelencia Internacional UAM+CSIC". F.M. would also like to acknowledge the Mainz Institute for Theoretical Physics (MITP) for its hospitality and support during the completion of this work.

\clearpage

\appendix

%%% Ap. A

\section{$\CN=1$ supersymmetric conditions and generalised Dolbeault operator}
\label{susy}

In this appendix we write down the $\mathcal N=1$ supersymmetry equations for a 4d Minkowski compactification of Type II string theories in terms of a generalised Dolbeault operator following \cite{Tomasiello:2007zq}. We also rederive the results of section \ref{s:IIA} in a way that can be applied to more general situations. We refer the reader to \cite{Tomasiello:2007zq,Martucci:2009sf,ty12} more more details. We follow the conventions in \cite{Martucci:2009sf}.

We consider the following ansatz for the metric in the 10 space $\IR^{1,3}\times X_6$,
\be
ds^2=e^{2A}ds^2_{\IR^{1,3}}+ds^2_{X_6}
\ee
where the warp factor depends generically on the coordinates $y^m$ on $X_6$. We take as independent degrees of freedom the RR field strengths with legs only on $X_6$, namely, $F_0,\,F_2$ and $F_4$, that we arrange into a polyform $F$. In the presence of D-branes and O-planes, this polyform satisfies the Bianchi identity
\be
dF=-j
\ee
with $j$ the corresponding current. More explicitly, for a single D-brane on $\Sigma_\a$ with gauge flux $F_\a$ we have $j_\a=\delta(\Sigma_\a)\wedge e^{-F_\a}$. Also, the $H$-field should satisfy its own Bianchi identity $dH=0$.

This kind of compactifications are completely determined by two complex polyforms $\mathcal Z$ and $T$ of opposite parity that define an $SU(3)\times SU(3)$-structure. Generically, only one of them is integrable, which we take to be $\mathcal Z$, and it defines an integrable generalised complex structure $\mathcal J$ on $X_6$.
The explicit expressions of the polyforms $\mathcal Z$ and $T$  for the cases of type IIA with O6-planes and type IIB with O3/O7-planes are
\be\begin{split}
\text{IIA}: & \qquad \mathcal{Z} = e^{3A-\phi}e^{i J+B}\,,\qquad T = e^{-\phi}\,\Omega \wedge e^B\,,\\
\text{IIB}: & \qquad \mathcal{Z} = e^{-\phi}\,\Omega \wedge e^B\,,\qquad T =  e^{3A-\phi}e^{i J+B}\,.
\end{split}\ee
Given these definitions, the supersymmetry conditions can be written as \cite{Tomasiello:2007zq,Martucci:2009sf}
\be\label{scond}
d\mathcal Z=0,\qquad d(e^{2A}\im\, T)=0,\qquad d\,\re \,T=-\mathcal J\cdot F.
\ee
Notice that none of these conditions involves the metric or B-field explicitly, which are encoded in the polyforms $\mathcal Z$ and $T$ that characterise the internal manifold.

In order to preserve supersymmetry, the (backreacted) sources must be calibrated \cite{k05,ms05}. This means that they should wrap generalised complex submanifolds, i.e. $\mathcal J\cdot j=0$ and also that $\langle \im\, T,j\rangle =0$. It has been shown \cite{kt07} that the supersymmetry equations \ref{scond} together with the calibration condition for the sources and the Bianchi identities imply that the whole set of equations of motion are satisfied. However, as stressed in the main text, one should also impose that the field strengths are quantised to have a consistent supersymmetric solution. 

In the following we shall give a different derivation of the results of section \ref{s:IIA}. The key point for this derivation is that, using the equations above, we can write the RR field strength as
\be\label{exact}
F=-\dj\,\re\, T
\ee
where we defined the operator $\dj=[d,\mathcal J]$ that satisfies $(\dj)^2=0$. This is the generalisation of $d^c=-i(\p-\bar \p)$ in complex geometry, where $\p$ is the Dolbeault differential. Thus, introducing $\dj$ is equivalent to defining a generalised Dolbeault operator. 

\subsection*{Symplectic cohomologies}

Here we arrive at the main result in section \ref{s:IIA} in terms of eq.(\ref{exact}) and of certain cohomology groups that can be defined in a symplectic manifold \cite{ty12} . This is useful since it shows that the reasoning is more general and only depends on the integrable generalised complex structure on $X_6$, which behaves nicely when deforming the sources. For Type IIA with D6-branes and O6-planes it corresponds to a symplectic structure so we will focus on this case.

We give the relevant definitions and results without proof since they can be found in \cite{ty12} (see also \cite{caval}). Let $(X_6,J)$ be a compact symplectic manifold, then the operator $\dj$ is given by $\delta=[d,J^{-1}\mathrel{\llcorner}]$ where $\mathrel{\llcorner}$ denotes index contraction. One can then define the following cohomology groups 
\be
H^k_{d+\delta}(X_6)=\frac{\text{ker}(d+\delta)\cap\Omega^k(X_6)}{\text{im}\,d\delta\cap\Omega^k(X_6)},\qquad H^k_{d\delta}(X_6)=\frac{\text{ker}(d\delta)\cap\Omega^k(X_6)}{(\text{im}\,d+\text{im}\,\delta)\cap\Omega^k(X_6)}
\ee
where $\Omega^k(X_6)$ is the space of $k$-forms on $X_6$. The elements in the first group are forms $\a_k$ such that $d\a_k=\delta\a_k=0$ and $\a_k\neq d\delta \beta_k$ for all $\beta_k$. Regarding the second group, it consists of forms $\a_k$ such that $d\delta \a_k=0$ and $\a_k\neq d\b_{k-1}$, $\a_k\neq\delta \beta_{k+1}$ for every $\b_{k-1}$ and $\b_{k+1}$. One can prove that there exists a non-degenerate pairing between $H^k_{d+\delta}(X_6)$ and $H^{6-k}_{d\delta}(X_6)$ given by
\be
(\a_k,\b_{6-k})=\int_{X_6}\a_k\wedge \b_{6-k}.
\ee
Furthermore, there is an isomorphism between $H^k_{d+\delta}(X_6)$ and $H^{6-k}_{d+\delta}(X_6)$ given by $(J\wedge)^{3-k}$. Finally, if $X_6$ satisfies the $d\delta$-lemma (which we assume),  there is yet another isomorphism between $H^k_{d+\delta}(X_6)$ and the usual de Rham cohomology $H^{k}_{dR}(X_6)$.

Now we have all the necessary ingredients to arrive at the result in the main text. Equation (\ref{exact}) says that $F_2$ (which is the only non-vanishing RR field strength) is the trivial class in $H^2_{d\delta}(X_6)$ since $d\delta F_2=0$ and $F_2=-\delta(\re\, T)$. Thus, given the pairing with $H^4_{d+\delta}(X_6)$, we may write
\be
\int_{X_6}F_2\wedge Q_4=0\qquad \forall\, Q_4\in H^4_{d+\delta}(X_6).
\ee
Using the isomorphisms quoted above we can rewrite this condition in terms of the de Rham cohomology, namely
\be
\int_{X_6}F_2\wedge J\wedge \omega_2=0\qquad \forall \,\omega_2\in H^2_{dR}(X_6).
\ee
Deforming the location of the D6-branes produces a $\Delta F_2$ that must satisfy the equation above to preserve supersymmetry, which is precisely the same as eq.(\ref{int6}) in the main text. The rest of the argument involving the quantisation condition of $F_2$ is the same so we do not repeat it here.

\section{Linear equivalence}
\label{Leq}

Linear equivalence is a criterion to compare different cycles that is finer than homology. It can be regarded as a generalisation to higher codimension of the more familiar case of divisors in complex geometry. In that case, one assigns a line bundle to every divisor and then compares such bundles. The same idea can be applied to higher codimension submanifolds by assigning a gerbe to them. See \cite{Hitchin99} for a nice mathematical explanation of this concept.

In this appendix we take a more pragmatical point of view and discuss its relevance when dealing with Bianchi identities with localised sources. Let $\mathcal M$ be a Riemannian manifold of dimension $n$ and take a collection of $(p-n)$-cycles $\{\Pi_\a\}$. Now, consider the following differential equation for the $(p-1)$-form $F_{p-1}$
\be
dF_{p-1}=\sum_\a \delta(\Pi_\a)
\ee
where $\delta(\Pi_\a)$ is a $p$-form with support on $\Pi_\a$ and all its legs normal to it. This is clearly the Bianchi identity for a field strength $F_{p-1}$ with localised (magnetic) sources. If we want to have a globally well-defined solution we must impose the tadpole condition, namely
\be\label{aptad}
\sum_\a\,[\Pi_\a]=0,
\ee
which we assume in the following. Furthermore, if $F_{p-1}$ corresponds to a field strength, it must be quantised. In other words, $F_{p-1}\in H^{p-1}(\mathcal M,\IZ) $ in some normalisation.

We can now use the Hodge decomposition to write the field strength as
\be
F_{p-1}=A+dB+d^*C
\ee
where $A$ is a smooth harmonic $(p-1)$-form on $\mathcal M$ and $d^*$ is the adjoint of $d$. Clearly, the coexact term $d^*C$ is fixed by the Bianchi identity but the exact piece $dB$ is not. If there are no electric sources for $F_{p-1}$ we have that $d^*F_{p-1}=0$ which forces such exact term to vanish (which we assume). 
Thus, only the harmonic contribution remains unfixed and, according to the results in \cite{Hitchin99}, it is the quantisation condition that does the job and where the notion of linear equivalence appears.

For our purposes, we may define linear equivalence as follows. We say that the collection of cycles $\{\Pi_\a\}$ is linearly equivalent to zero if, given the setup described above, the harmonic term is an integral form, i.e. $A\in H^{p-1}(\mathcal M,\IZ)$. This means that, even if we take $A=0$, the field strength will be quantised.

It is not easy to check in a particular example whether a set of cycles are linearly equivalent to each other. However, there is an alternative characterisation that is somewhat more practical. Namely, we have that $\{\Pi_\a\}$ are linearly equivalent to zero if 
\be
\int_\Sigma\omega\in\IZ 
\ee
for each harmonic $(n-p+1)$-form $\omega$ in $\mathcal M$, where $\Sigma$ is a $(n-p+1)$-chain such that $\p\Sigma= \sum_\a\Pi_\a$ (see  \cite{Hitchin99} for a proof). The generalisation of linear equivalence that applies to D-branes (seen as submanifolds with bundles on them) has been worked out in \cite{Marchesano:2014bia}.

\end{document}